\documentclass{aa}  
\usepackage{graphicx}
\usepackage{txfonts}
\usepackage{natbib}
\usepackage{aalongtable}
\newcommand{\Msun}{~M_\odot} 
\newcommand{\lsim}{\raise0.3ex\hbox{$<$}\kern-0.75em{\lower0.65ex\hbox{$\sim$}}}
\newcommand{\gsim}{\raise0.3ex\hbox{$>$}\kern-0.75em{\lower0.65ex\hbox{$\sim$}}}

\begin{document}

\title{Supernova 2006aj and the associated X-Ray Flash 060218\thanks{\rm This paper is based on observations from the ESO/Danish 1.5-m telescope at the
La Silla Observatory  and on 
observations made with the Nordic Optical Telescope, operated
on the island of La Palma jointly by Denmark, Finland, Iceland,
Norway, and Sweden, in the Spanish Observatorio del Roque de los
Muchachos of the Instituto de Astrofisica de Canarias.
}}

\author{
J.~Sollerman\inst{1,2}
\and
A.~O.~Jaunsen\inst{3}
\and
J.~P.~U.~Fynbo\inst{1}
\and
J.~Hjorth\inst{1}
\and
P.~Jakobsson\inst{1}
\and
M.~Stritzinger\inst{1}
\and
C.~F\'eron\inst{1}
\and
P.~Laursen\inst{1}     
\and
J.-E.~Ovaldsen\inst{3}
\and
J.~Selj\inst{3}
\and
C.~C.~Th\"one\inst{1}
\and
D.~Xu\inst{1}
\and
T.~Davis\inst{1}
\and
J.~Gorosabel\inst{4}
\and
D.~Watson\inst{1}
\and
R.~Duro\inst{3}
\and
I.~Ilyin\inst{5}
\and
B.~L.~Jensen\inst{1}
\and
N.~Lysfjord\inst{3}
\and
T.~Marquart\inst{6}
\and
T.~B.~Nielsen\inst{7}
\and
J.~N\"ar\"anen\inst{8}
\and
H.~E.~Schwarz\inst{9}
\and
S.~Walch\inst{10}
\and
M.~Wold\inst{11}
\and
G.~\"Ostlin\inst{2}
}
\institute{
Dark Cosmology Centre, Niels Bohr Institute, University 
of Copenhagen, Juliane Maries Vej 30, DK--2100 Copenhagen~\O, Denmark
\and
Stockholm Observatory, Department of Astronomy, AlbaNova, 
S-106 91 Stockholm, Sweden
\and
Institute of theoretical astrophysics, PO Box 1029, N-0315 Oslo, Norway
\and
Instituto de Astrofisica de Andalucia (IAA-CSIC), PO Box 03004, 18080 Granada, Spain
\and
Astrophysikalisches Institut, An der Sternwarte 16, 14482 Potsdam, Germany
\and
Dept. of Astronomy and Space Physics, Box 515, S-751 20 Uppsala, Sweden
\and
NOT, Apartado 474, E-38700 Santa Cruz de la Palma, Spain
\and
Observatory, University of Helsinki, P.O. Box 14, FIN-00014, Finland
\and
CTIO-NOAO, Casilla 603, La Serena, Chile
\and
University Observatory Munich, Scheinerstr. 1, 81679 Munich, Germany
\and
ESO, Karl-Schwarzschild-Strasse 2, 85744 Garching, Germany 
}

   \date{}

% \abstract{}{}{}{}{} 
% 5 {} token are mandatory
 
\begin{abstract}
{}
{We have studied the afterglow of
the gamma-ray burst (GRB) of February 18, 2006. 
This is a nearby long GRB, with a very low peak energy, and is therefore classified as an X-ray Flash (XRF).
XRF\,060218 is clearly associated with a supernova -- dubbed SN\,2006aj.}
{We present early spectra for SN\,2006aj as well as optical lightcurves reaching out to 50 days past explosion.} 
{Our optical lightcurves define the rise times, the lightcurve shapes and the absolute magnitudes in the 
$U$, $V$ and $R$ bands, and we compare these data with data for other relevant supernovae. SN\,2006aj evolved quite fast, somewhat similarly to SN\,2002ap, but not as fast as SN\,1994I.
Our spectra show the evolution of the supernova over the peak, when the $U$-band portion of the spectrum rapidly fades due to extensive line blanketing.
We compare to similar spectra of very energetic Type Ic supernovae. Our first spectra are earlier than spectra for any other GRB-SN. The spectrum taken 12 days after burst in the rest frame is similar to somewhat later spectra of both SN\,1998bw and SN\,2003dh, implying a rapid early evolution. This is consistent with the fast lightcurve.

From the narrow emission lines from the host galaxy we derive a redshift of 
$z=0.0331\pm0.0007$. This makes XRF\,060218 the second closest gamma-ray burst detected. The flux of these emission lines indicate a high-excitation 
state, and a modest metallicity and star formation rate of the host galaxy.} 
{}
\end{abstract}


   \maketitle

\section{INTRODUCTION}\label{introduction}

The last few years have settled the debate about the origin of long gamma-ray bursts (GRBs). The hint provided by GRB\,980425 and SN\,1998bw \citep{galama98} was finally taken when the spectroscopic follow-up of the afterglow of GRB\,030329 revealed the unambiguous signatures of a very energetic supernova -- SN 2003dh \citep{hjorth03,matheson03,stanek03}. Soon thereafter, another clear-cut SN\,1998bw look-alike emerged in the afterglow of GRB\,031203 \citep{malesani04,thomsen04}. 
While the Swift satellite \citep{gehrels04} has been very successful in finding GRBs over a large redshift range \citep[e.g.,][]{jakobsson06}, the wait for the next spectacular case of a nearby 
GRB-supernova has lasted more than two years.

\subsection{GRB\,060218}

GRB\,060218 was detected by the BAT instrument on-board the Swift satellite 
\citep[][]{cusumanogcn4775} on February 
18.149 2006 UT.
% (Universal Time is used throughout the article).
This burst had exceptional high-energy properties \citep{campana06}. 
The peak energy of the event (Sect~\ref{xrays}) was very low and we will hereafter refer to this burst as an X-ray flash.
XRF\,060218 is 
one of the longest bursts ever detected, and the unusual properties gave a very confused early impression. Several GCNs indicated that this was probably not a proper GRB, and our optical monitoring programme was therefore somewhat delayed. 
However, eventually this turned out to be a very interesting low-$z$ event \citep{mirabalgcn4792} with a likely association to a supernova \citep{masettigcn4803,soderbergiau}.
The transient has now been detected over a wide wavelength range, from X-rays \citep{kenneagcn4776} to radio \citep{soderberggcn4794}. 

In this paper we focus on the optical transient, and the early spectral and photometric evolution of this supernova (SN). The paper is organized as follows. 
Sect.~\ref{observations} outlines how the optical observations were obtained and reduced. The results are presented in Sect.~\ref{results}, which includes $U$-, $V$- and $R$-band lightcurves as well as spectra of the SN, and an analysis of the host galaxy.
We end the paper with a discussion (Sect.~\ref{discussion}) where we compare the properties of this SN with other relevant SNe.

\section{OBSERVATIONS}\label{observations}

\subsection{Photometry}\label{photobservations}

The observations for XRF\,060218 were somewhat complicated and hampered by the celestial position of the burst. Being close to the Sun it could only be observed for a short time right after sunset. We have used the combined efforts of two telescopes, at a northern and a southern observatory, to follow the object until it faded into the glare of the Sun, about 50 days past the burst. For the final observations we had to restrict ourselves to a single passband ($R$) due to the limited time available for observations in the twillight.

We obtained imaging of the transient of XRF\,060218 with the
ESO/Danish 1.5~m telescope (D1.5m) on La Silla 
equipped with the DFOSC instrument, which offers a $13.7\times13.7$ arcminute 
field-of-view (FOV) at 0.395 arcsec per pixel. 
We also used the 2.56~m Nordic Optical Telescope (NOT) on La Palma equipped with ALFOSC which offers a FOV of $6.3\times6.3$ arcminutes with a pixel scale of 0.189 arcsec, as well as StanCam which has a pixel scale of 0.176 arcsec over $3\times3$ arcminutes.

The journal of observations is given in Table~\ref{t:log}.
The data were reduced using standard techniques for de-biasing and
flat-fielding.

\subsection{Spectroscopy}\label{specobservations}

Spectra of the source were obtained with ALFOSC at four epochs, February 21, 22 and 24 and on March 2. These epochs correspond to 
3.78, 4.71, 6.71 and 12.71 days past the burst.
Each spectrum had an integration time of 2400 
seconds using grism 4 and a 1.3 arcsec wide 
slit. This set-up provided a dispersion of 3~\AA~per pixel.
The spectral range covered is from 3300 to 9200 \AA. There is some second order contamination above 6600~\AA, and significant fringing above $\sim7500$~\AA.
The spectra were taken at or close to the parallactic angle. 
We note that apart from the first spectrum, taken at an airmass of 1.93, 
all spectra were achieved at an airmass $<1.5$. The NOT/ALFOSC has a
high efficiency
in the UV, 
so we put emphasis on obtaining the bluest part of the spectrum.

The spectra were reduced following standard procedures in {\tt MIDAS}
and {\tt IRAF}. Wavelength calibration was achieved by comparison to images taken of helium and neon lamps. The flux calibration was performed using the spectrophotometric standard star GD71 \citep{bohlin95}, which was observed every night close in time to the supernova observation.
Finally, the absolute flux-calibration was achieved by comparison to the contemporary (or interpolated) $V$-band photometry.

When comparing to synthetic photometry obtained by integrating each spectrum under the filter profiles, we discovered that some of the spectra have suffered from differential slit losses. This has been considered in the analysis below.

\section{RESULTS}\label{results}

\subsection{The Lightcurves}\label{lightcurves}

Aperture photometry of the transient was carried out using
a combination of {\tt DAOPHOT} \citep{stetson87} and {\tt SExtractor}
\citep{bertin96}.
We measured the magnitudes of the supernova as well as for 9
stars in the field in the $V$ and $R$ bands (7 local standards in the $U$ band). The relative magnitudes were transformed to the
standard system
using observations of photometric standard stars \citep
{landolt}.
We estimate an absolute photometric accuracy of 0.08, 0.06 and 0.04  
mags in the $U$, $V$ and $R$ bands, respectively.

\begin{figure*}
{\includegraphics[width=\textwidth,angle=0,clip]{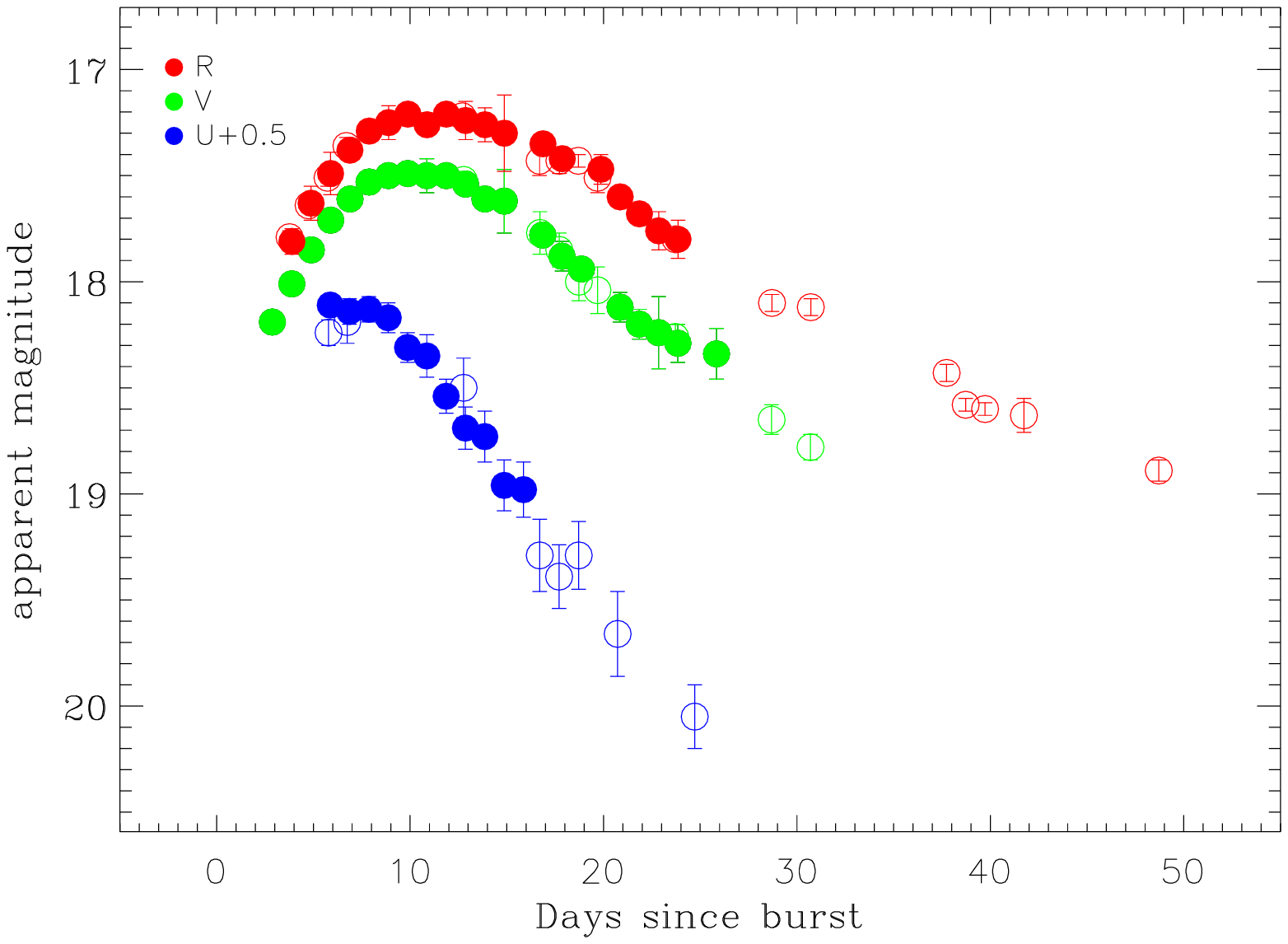}}
\caption{}{The $U$-,  $V$- and  $R$-band lightcurves of SN\,2006aj.
Dates are given in days after the high-energy burst in the observers frame.
The filled circles are data from the D1.5m, and the open circles from the NOT
(data from Table~\ref{t:log}). For clarity, we have excluded points for which the errors are greater than 0.15 mag when more accurate data were available for the same night.
These magnitudes are not corrected for the host galaxy contribution, and have not been K-corrected. Corrections for the host galaxy is done in Fig.~\ref{f:lightcurvescompare} 
}
\label{f:lightcurves}
%\end{flushleft}
\end{figure*}

In Fig.~\ref{f:lightcurves} we plot the $U$-, $V$- and $R$-band lightcurves. 
This is the data from Table~\ref{t:log}.
The dates are given with respect to the time of the 
burst in the observers frame. We have not plotted the data with errors larger than 0.15 mag, if there are more accurate data from the same night.
The $R$- and $V$-band lightcurves are followed from well before peak and are 
traced to way past maximum. 

In Fig.~\ref{f:lightcurves} we have applied no corrections 
to subtract the host galaxy (estimated at $R=19.9$, see Sect.~\ref{host}). 
This can be an important contribution, $\sim0.1$~mag 
at maximum light, and is considered in the following analysis. 
Also, we have made no K-corrections for the magnitudes given in Table~\ref{t:log} and plotted in Fig.~\ref{f:lightcurves}. At the early epochs where we have spectra, we estimate this correction to be 
$\sim0.04$~mag in the $V$ band, and $0.11$~mag in the $R$ band. The final spectrum is taken closest in time to the maximum light in these bands, and indicate K-corrections of $\sim0.02$~mag in the $V$ band, and $0.15$~mag in $R$.

To estimate the time of maximum, peak brightness and the lightcurve shape as described by $\Delta$m$_{15}$ (the number of magnitudes the supernova decayed in the 15 days following maximum brightness) we have fitted the lightcurves with smooth functions \citep[see][]{stritzinger05}. 
We estimate the rise times of $t(V)=10.4\pm0.5$ days past burst and 
$t(R)=11.4\pm0.5$ days. 
We further estimate  $\Delta$m$_{15}(V)=0.92$ and $\Delta$m$_{15}(R)=0.71$~mag from the observed data. When correcting for the underlying emission from the host galaxy (Sect.~\ref{other}), as well as for time dilation, the corrected numbers are 
$\Delta$m$_{15}(V)=1.1\pm0.1$ and $\Delta$m$_{15}(R)=0.90\pm0.1$~mag.

The peak magnitudes are estimated to be m($V$)=$17.47\pm0.05$ and 
m($R$)=$17.22\pm0.05$~mag. To determine the absolute magnitudes we need estimates of the distance and extinction.
The redshift of this burst is $z=0.0331$
\citep[][ see also Sect.~\ref{spectrum} ]{mirabalgcn4792}, 
and assuming a cosmology 
where $H_0=70$\,km\,s$^{-1}$\,Mpc$^{-1}$, $\Omega_\Lambda = 0.7$
and $\Omega_{\rm m}=0.3$,
this corresponds to a luminosity distance of 145.4 Mpc.

The reddening associated with 
Galactic extinction is E($B-V$)=0.14~mag according to the maps by \citet{schlegel98}. High-resolution spectra \citep{guenthergcn4863} can be used to check this. Using the sodium lines to estimate the reddening \citep[see e.g.,][]{munari97} provides E($B-V$)=0.127 for the Milky Way, and E($B-V$)=0.042~mag for the GRB host. This is consistent with adopting a total reddening of E($B-V$)=0.14~mag. We note that \citet{campana06} required E($B-V)_{\rm host}=0.20$ mag based on the assumption of thermal radiation detected by UVOT. This is higher than claimed by \citet{guenthergcn4863}. It is known that the Na~I~D lines do not provide a robust measure of the extinction, and could be influenced by e.g., the ionization state in the host \citep[see e.g.,][]{sollerman05a,sollerman05b}. However, the overall properties of the host galaxy based on the spectral energy distribution (SED) modeling (Sect.~\ref{modeling}), as well as the measured Balmer line decrement \citep[Sect.~\ref{other}, see also][]{pian06}, also argue for a low host extinction. In the following we will therefore adopt a total extinction of E$(B-V)=0.14$~mag.

The absolute magnitudes of the SN are then M$(V)=-18.8$ and M$(R)=-18.9$ mag. Finally, correcting these estimates for host contamination 
($V,R=0.09,0.10$) and K-corrections ($V,R=0.02,0.15$)
our best estimates are M$(V)=-18.7$ and  M$(R)=-18.7$ mag. These are the magnitudes adopted for comparison to other SNe, and are given in Table~\ref{t:summary}.

\subsubsection{The $U$-band lightcurve}\label{ubandlc}

Lightcurves for Type Ic SNe are relatively rare in the $U$ band. For SN\,2006aj, we started our $U$-band imaging campaign 5 days past the burst. At this epoch, the $U$ band was already close to maximum light. We then followed the evolution of the $U$-band flux until 25 days past the burst, after which the supernova became to faint (also compared to the host) to allow further monitoring.

Given the sparse pre-maximum coverage, the estimates are somewhat more uncertain in the $U$ band. We estimate $t(U)=6.8\pm1.0$ days. 
The estimate of the corrected light curve shape 
is rather uncertain, due to the large correction for host contamination 
on the already steep lightcurve. 
We estimate $\Delta$m$_{15}(U)=2.0\pm0.2$.
The peak brightness is m$(U)=17.60\pm0.10$ mag, which converts to an absolute $U$-band magnitude of M$(U)=-18.9$ in the Vega magnitude system.

The K-corrections are most uncertain in the $U$ band, and could also be significant in particular for the latest epochs where the spectrum falls very steeply in that region. At around $U$-band maximum light, we estimate a K-correction of $\sim-0.15$~mag. Applying this K-correction, and a correction for the host galaxy ($0.08$~mag) we therefore estimate the final absolute magnitude M($U)=-18.9$.

We summarize all corrected lightcurve parameters in Table~\ref{t:summary}. In this table, the rise times are corrected for time dilation, as are the light curve shapes which are also corrected for the underlying host galaxy. The absolute magnitudes are corrected for extinction, host galaxy contamination and are K-corrected.

\subsection{The Spectral Evolution}\label{spectrum}

The flux-calibrated spectra are shown in Fig.~\ref{f:spectra}. 
The continuum-like spectrum with broad bumps renders a classification of this burst as a Type Ic supernova, based on the lack of conspicuous SN lines \citep[e.g.,][]{patat01}.

The spectral evolution is well represented, the most obvious development being the depression of the UV flux with time.
Some of the depression seen in our final spectrum may be attributed to differential slit losses, but the overall evolution of the spectra are correct, as can be seen from the comparison to the broad band light curves.
 In fact, the UV depression is a common feature of SNe and reflects the increased line blanketing due to low ionization iron group elements. As seen from the lightcurve, the $U$ band actually peaked close to the date of our third spectrum, so we see the rise of the $U$ band up to that epoch in the spectral evolution, followed by a rapid decline to the final spectrum. 

Our latest spectrum, taken 12.3 days past burst in the rest frame, shows a dramatic evolution of the flux towards the red part of the spectrum. The broad red bumps are common features of so-called hypernovae and signal huge expansion velocities of the ejecta. Interpreting the inflection point at $\sim6080~$\AA~as the Si II 6355 \AA~feature seen in other GRB-SNe \citep{patat01,hjorth03}, we can estimate an expansion velocity of $\sim 22\,000$~km~s$^{-1}$. At 12 days past burst, this is similar to the expansion velocities measured in SN\,1998bw and SN\,2003dh. However, since this feature is quite loosely defined, this estimate can only be approximate.

From the multitude of narrow emission lines from the host galaxy
we can also measure the redshift to the supernova.
The positions and fluxes of a number of detected narrow  
lines are given in Table~\ref{t:lines}. We derived the redshift by
measuring the positions of the
%[\ion{O}{2}] line, the [\ion{O}{3}] lines as well as 
[O~II] line, the [O~III] lines as well as 
H$\alpha$ and H$\beta$ at all 4 epochs, and conclude $z=0.0331\pm0.0007$. 

The fluxes of the lines were measured by Gaussian fits, using both {\tt IDL} and {\tt IRAF splot}. We use these below to estimate the star formation rate and the metallicity. We note that the values given in Table~\ref{t:lines} are averages for the four spectra
corrected for E$(B-V)=0.14$~mag. 
Apart from the stronger lines listed in Table~\ref{t:lines}, we also detect 
[Ne~III] $\lambda3869$, which signals the presence of ionizing radiation. 
This line is about 3 times weaker than H$\beta$, although the uncertainty in such a weak line is $\sim50\%$ in our spectra. The stronger lines have uncertainties of $\lesssim20\%$.

\subsection{The Host Galaxy}\label{host}

\subsubsection{Modeling the SED}\label{modeling}

The $ugr{i}z$-band SDSS
pre-imaging of the field \citep{coolgcn4777} 
allowed us to construct the optical spectral 
energy distribution of the host galaxy. 
However, it was noted \citep{hicken06gcn4898,modjaz06} that the absolute calibration of this field was not correct.
To correct the SDSS model magnitudes we used field star photometry 
\citep{hicken06gcn4898}, which was transformed from Landolt to the SDSS-system using \citet{jester05}. 
Corrected host magnitudes were then deduced from the offsets between the
SDSS model magnitudes and the transformed values, giving:
$u=21.24\pm0.15,  g=20.29\pm0.04,  r=20.16\pm0.03,  i=19.96\pm0.04,
z=19.80\pm0.13$.
We then used the best fit SED (see below) to transform these magnitudes back 
to Landolt photometry:
%Applying the inverse transformation from \citet{jester05} on these
%corrected host
%magnitudes, we get an estimate of the Landolt host photometry:
$U = 20.45\pm0.15, B = 20.46\pm0.07, 
V = 20.19\pm0.04, R = 19.86 \pm0.03, I = 19.47\pm0.06~$mag.
These are the host galaxy magnitudes used to correct the light curve parameters in Sect.~\ref{lightcurves}.

The $ugr{i}z$ host galaxy photometric points were then
de-reddened by 
the Galactic extinction following \citet{schlegel98}
and then
fitted based on the 
SDSS filter+CCD efficiency  curves \citep{fukugita96}
and using the synthetic SED  templates constructed with the HyperZ code  
\citep{bolzonella00}.

For  the construction  of the  synthetic templates  three  initial mass 
functions (IMFs) were used 
\citep{scalo86,miller79,salpeter55}.
We also used four different extinction laws:
MW \citep{seaton79}, LMC \citep[][]{fitzpatrick86}, SMC \citep{prevot84}
and one for starburst galaxies \citep{calzetti00}.
Solar metallicity was assumed for all the templates. 
The  redshift of the templates was  fixed at $z=0.0331$.  
In  addition, a  wide range of  star-formation  histories were
considered 
\citep[see more  details on the $\tau$ parameter in][]{gorosabel05},
creating  different families of  templates:
Elliptical, Starburst, Lenticular, Irregular and Spiral galaxies.

The  $ugr{i}z$-band
photometric  points  were  satisfactorily  fitted by  the SED templates
($\chi^2_{\rm d.o.f} \sim  1.3$; see  Fig.~\ref{f:SED}).
Our  SED fits did not favour any IMF, extinction law or galaxy type.
This means that the inferred host galaxy extinction is independent on the input model, and is stable at around $A_{\rm  V} =  
0.1 - 0.3$~mag. This is why we were favoring a low host galaxy extinction in Sect.~\ref{lightcurves}.

\subsubsection{Host galaxy properties}\label{other}

The host magnitude of $B=20.46$~mag, 
and the Galactic extinction of $E(B-V)=0.14$
gives an absolute magnitude for the host of 
M$(B)=-15.9$ mag at the measured redshift. 
Adopting $M^{\rm B}_{\star}=-21.1$ this corresponds to 
$L=0.008L^{\rm B}_{\star}$.

From the H$\alpha$ and [O~II] lines we can 
estimate the star formation rate (SFR).
From both these lines we get
SFR $\sim0.05\Msun$~yr$^{-1}$, following \cite{kennicutt98}. This is of course only measured from the part of the galaxy that falls on the spectroscopic slit.
The specific star formation rate for the host galaxy of XRF\,060218 
is thus $\sim$6~M$_{\odot}$ yr$^{-1} (L/L^{\star})^{-1}$.

Finally, we can estimate the metallicity of the galaxy using the R23 technique. From the results presented in Table~\ref{t:lines}, we derive a log(R$_{23})=0.8-0.9$. This indicates a somewhat sub-solar metallicity, although the exact value can not be determined from this ratio alone \citep[see e.g., Fig.~5 by][]{kewley02}.

The luminosity and star formation rate thus indicates a small but fairly normal dwarf galaxy, similar to other nearby GRB host galaxies \citep{sollerman05b}. The low metallicity is also similar to that of other GRB host galaxies.

\section{DISCUSSION}\label{discussion}

\subsection{The supernova spectral evolution}

The spectral evolution reveals a rapidly evolving Type Ic supernova with very broad lines. Very few Type Ic spectra exist for such early epochs. Our first spectrum was obtained 3.7 days past the burst. The first spectrum for SN\,1998bw was not obtained until after a week, and for SN\,2003dh the emission was still dominated by the afterglow at this epoch. Therefore, it is difficult to make any one-to-one comparisons of the apparent bumps in these spectra with those of other similar SNe \citep[see e.g., Fig.~1 in ][]{mazzali02}.

In Fig.~\ref{f:comparison} we have re-plotted our latest spectrum of SN\,2006aj from March 2. This is 12.3 days past burst in the SN rest frame. We have also plotted spectra for SN\,1998bw \citep{patat01} and SN\,2003dh \citep{hjorth03}. These spectra are very similar to the one for SN\,2006aj, but are taken at a later epoch. SN\,2006aj thus displays a fast spectral evolution. This agrees with the narrow lightcurve.

\subsection{The Supernova Lightcurves}

From the lightcurves, as well as from the spectral evolution, we can see that the emission is dominated by the supernova rather than by the afterglow from very early on. This is similar to SN\,1998bw, where no optical afterglow was ever detected, but very different from SN\,2003dh which was dominated by the afterglow for more than a week before it emerged.

\subsubsection{The peak magnitude}

The peak magnitudes we have estimated show that SN\,2006aj was a fairly normal Type Ic supernova in that respect \citep{richardson06}. In particular, it was not as bright as  SN\,1998bw or SN\,2003dh. 
SN\,1998bw ejected $0.35-0.50\Msun$ of radioactive $^{56}$Ni \citep[see e.g., ][]{sollerman00,woosley99}. That SN\,2006aj was only $\sim50\%$ as luminous as SN\,1998bw
thus means that SN\,2006aj ejected $\sim0.22\pm0.06\Msun$ of radioactive $^{56}$Ni. This is still more than seen in other broad-line supernovae, such as SNe 1997ef and 2002ap. We note that the assumption that the peak magnitude scales with the nickel mass may not be valid for very asymmetric explosions \citep{hoeflich99}. GRBs are expected to be asymmetric, although they should all be pointed within a few degrees 
to our line of sight.

\subsubsection{The light curve shape}

The shape of the lightcurve is also of interest. 
For SN\,2006aj we have summarized the properties in Table~\ref{t:summary}. 
For comparison, the Type Ic SN\,1994I displayed 
$\Delta$m$_{15}(U)\sim2.5$ and
$\Delta$m$_{15}(V)\sim1.7$~mag.
The peak magnitude for SN\,1994I was reached after $\sim8$ days in $U$, 
and after 10 days in the $V$ band \citep{richmond96}. 
The rise time is, however, very uncertain for SN\,1994I; since the exact epoch 
of the explosion was not observed.

For SN\,1998bw, \cite{fynbo04} estimated
$\Delta$m$_{15}(U)\sim1.3$ and
$\Delta$m$_{15}(V)\sim0.7$~mag.
The peak magnitude was reached after 13.5 days in $U$, and after 17 days in the $V$ band \citep{galama98}. This is clearly slower than observed for SN\,2006aj.
Finally, SN\,2002ap reached $U$-band maximum at about 6.2 days \citep{foley03,galyam02,pandey03}. This lightcurve seems to be most similar to SN\,2006aj in this respect. In Fig.~\ref{f:lightcurvescompare} we compare the light curves of SN\,2006aj with those for SNe 1994I, 1998bw and 2002ap.
Note that in this figure we have corrected the light curves for time dilation (for SNe\,2006aj and 1998bw) and also corrected SN\,2006aj for the underlying host galaxy. This correction is quite substantial, in particular at late stages (compare Fig.~\ref{f:lightcurves}). The comparison in Fig.~\ref{f:lightcurvescompare} demonstrates that SN\,2006aj is in fact a fast version of SN\,2002ap.

Another important aspect of GRB-SNe is clearly the possibility to relate the supernova shock-wave breakout with the exact time of the explosion. \cite{campana06} used the UVOT instrument onboard Swift to follow the UV lightcurves from the early shock break-out to the following peak due to radioactive heating (the latter being the optical peak we are probing in this paper). Such a shock break-out 
 was also seen in SN\,1999ex \citep{stritzinger02} and in SN\,1998bw \citep{galama98}. For SN\,1999ex, the time of shock break-out could be estimated, and the rise time in the $V$ band was $t(V)=17.6$ days \citep{stritzinger02}. SN\,2006aj has a substantially faster lightcurve, which is related to the faster expansion velocities, and possibly also to a lower ejecta mass.

\subsubsection{X-ray Flash 060218}\label{xrays}

We note that GRB\,060218 was a very soft burst, and thus qualifies as an (unusual) XRF. \citet{campana06} estimated 
E$_{\rm peak}=4.9^{+0.4}_{-0.3}$~keV, at the very end of the observed distribution of peak energies. While the case for an association between long GRBs and SNe has been established (see Sect.~\ref{introduction}), the case is more unclear for XRFs.

XRF\,030723 showed a very conspicuous light curve bump
at $\sim16$ days past burst, suggesting 
the presence of a fast rising
supernova \citep{fynbo04}. 
In fact, doubts were raised against this interpretation since the required supernova light curve was very fast and narrow. The very fast $U$-band lightcurve of SN\,2006aj may be taken as support for the hypothesis of a SN in XRF\,030723. At a cosmological redshift of $\lesssim1$ the $R$-band light curve bump would correspond to rest frame $U$, as also noted by \cite{fynbo04}.

More recently, XRF\,050824 showed a less conspicuous bump \citep{sollerman06}. 
Moreover, XRF\,020903 has a lightcurve and spectrum consistent
with a supernova at $z=0.21$ \citep{soderberg05,bersier06}.
These findings all argue for a common progenitor for GRBs and XRFs.
The situation appeared less clear as other XRFs
with late-time coverage did not show clear evidence for a bright supernova bump
\citep[e.g.,][]{soderberg05}. 
However, with XRF\,060218 the case for a supernova origin for such bursts is obvious. 

Among the many remaining questions are the lack of conspicuous afterglow emission. Compared to GRB\,030329, the supernova emerged much faster from the afterglow for XRF\,060218. It is interesting to note that there was also no conspicuous afterglow in SN\,1998bw. For SN\,2003lw (GRB\,031203) there were claims of a very faint and fast decaying afterglow \citep{malesani04}. These bursts also had low values of E$_{\rm peak}$, and in fact \citet{watson04} considered 031203 to be an XRF. \cite{ramirez05} considered an off-axis model for XRF\,031203 in which this was really a normal GRB although viewed from an angle of about twice the opening angle.

\begin{figure*}
%\begin{flushleft}
{\includegraphics[width=\textwidth,angle=0,clip]{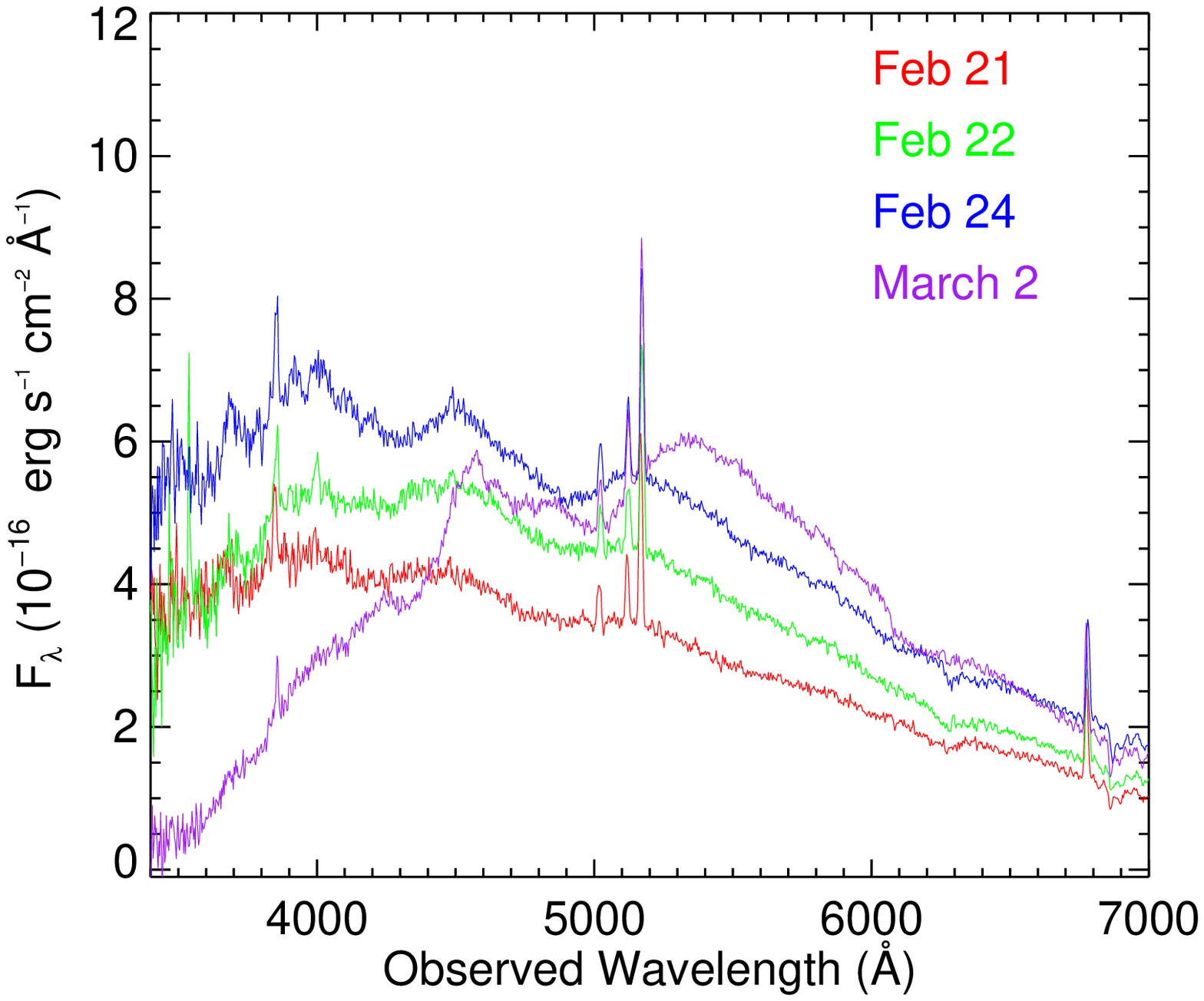}}
\caption{}{
The flux-calibrated and de-reddened spectra of the emerging supernova.
The spectra have been absolute flux-calibrated by comparison to $V$-band photometry.
In the Feb. 24 spectrum we note some components of intermediate widths $\sim3000-4000$~km~s$^{-1}$ in the blue part of the spectrum.
}
\label{f:spectra}
%\end{flushleft}
\end{figure*}

\begin{figure*}
\begin{flushleft}
{\includegraphics[width=0.7\textwidth,angle=-90,clip]{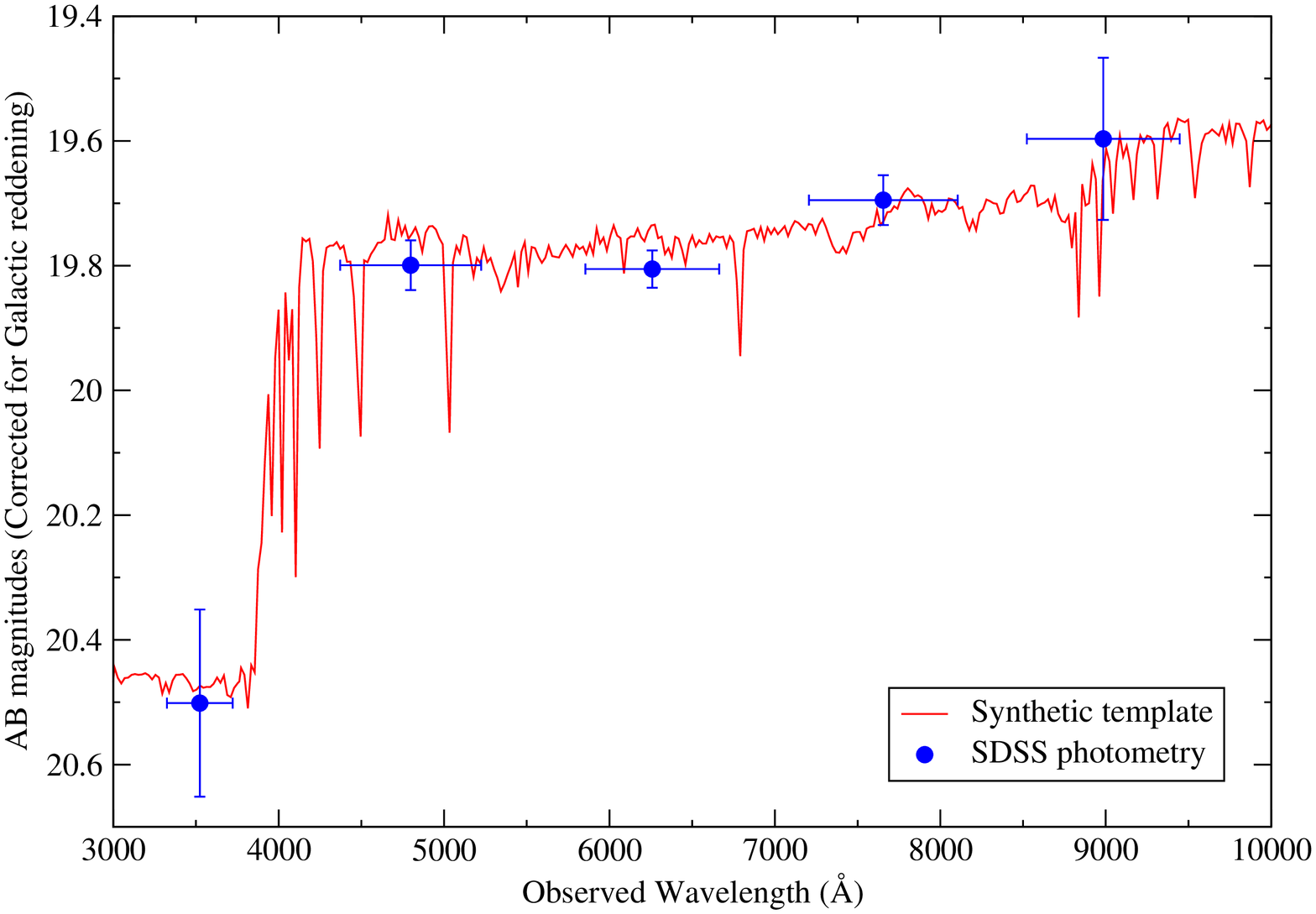}}
\caption{}{
The $ugr{i}z$-band SED 
of the XRF\,060218 host galaxy. The fit shows the best SED fit  
achieved ($\chi^2/d.o.f = 1.38$) when a SMC-like extinction law is 
assumed. This gives $A_{\rm V}=0.0$~mag. 
%and a dominant stellar population age of 53 Myr.
}
\label{f:SED}
\end{flushleft}
\end{figure*}

\begin{figure*}[h]
\begin{flushleft}
{\includegraphics[width=\textwidth,angle=0,clip]{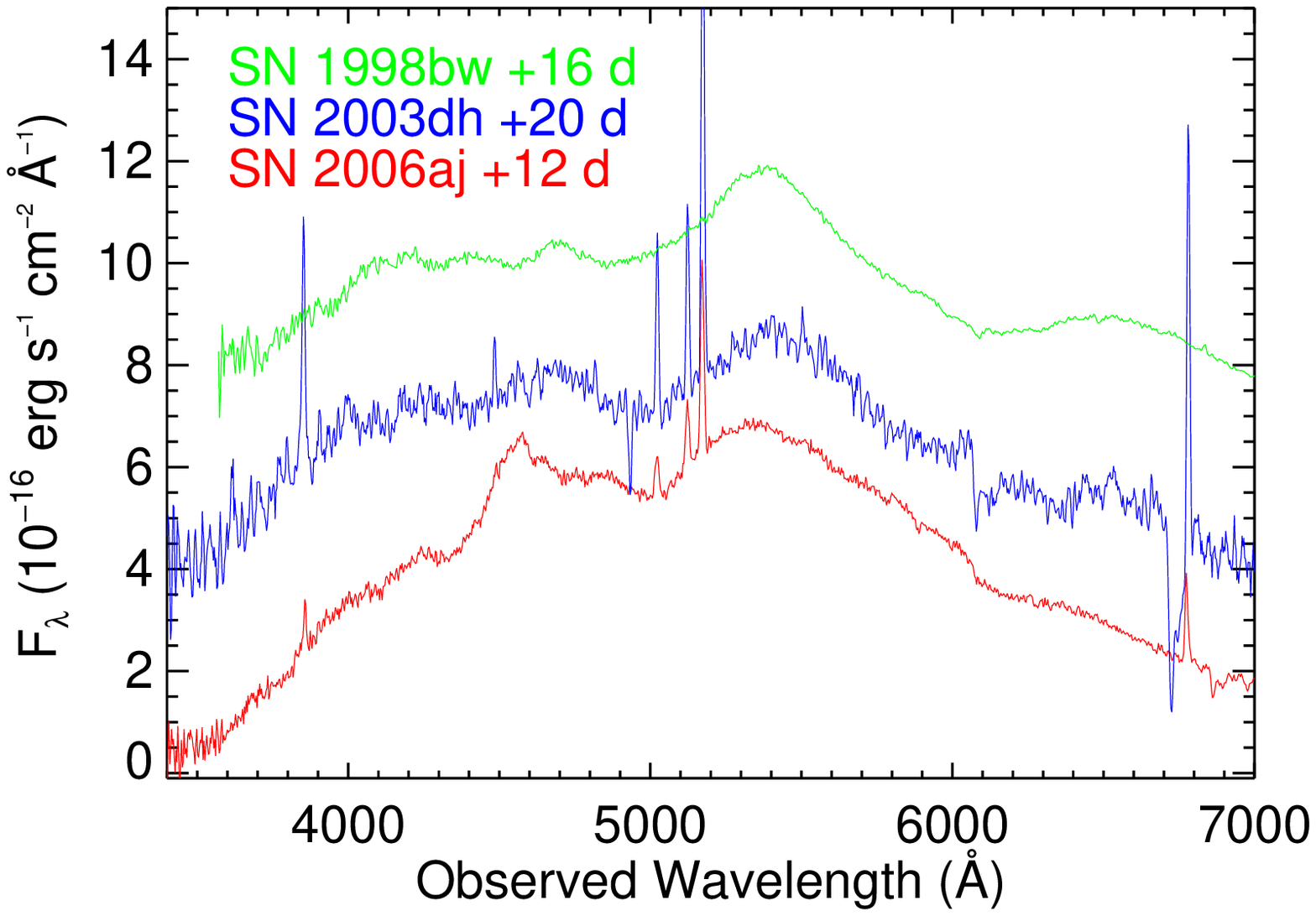}}
\caption{}{
Spectral comparison to other GRB SNe, SN\,1998bw at 16 days past burst and SN\,2003dh at 20 days past burst. Times are in the SN rest frames.
The spectra of SNe 1998bw and 2003dh have been arbitrarily shifted in flux.
They have also been shifted to the redshift of SN\,2006aj.
Note that SN\,2003dh has an afterglow that adds to the UV part.
}
\label{f:comparison}
\end{flushleft}
\end{figure*}

\begin{figure*}[h]
\begin{flushleft}
{\includegraphics[width=\textwidth,angle=0,clip]{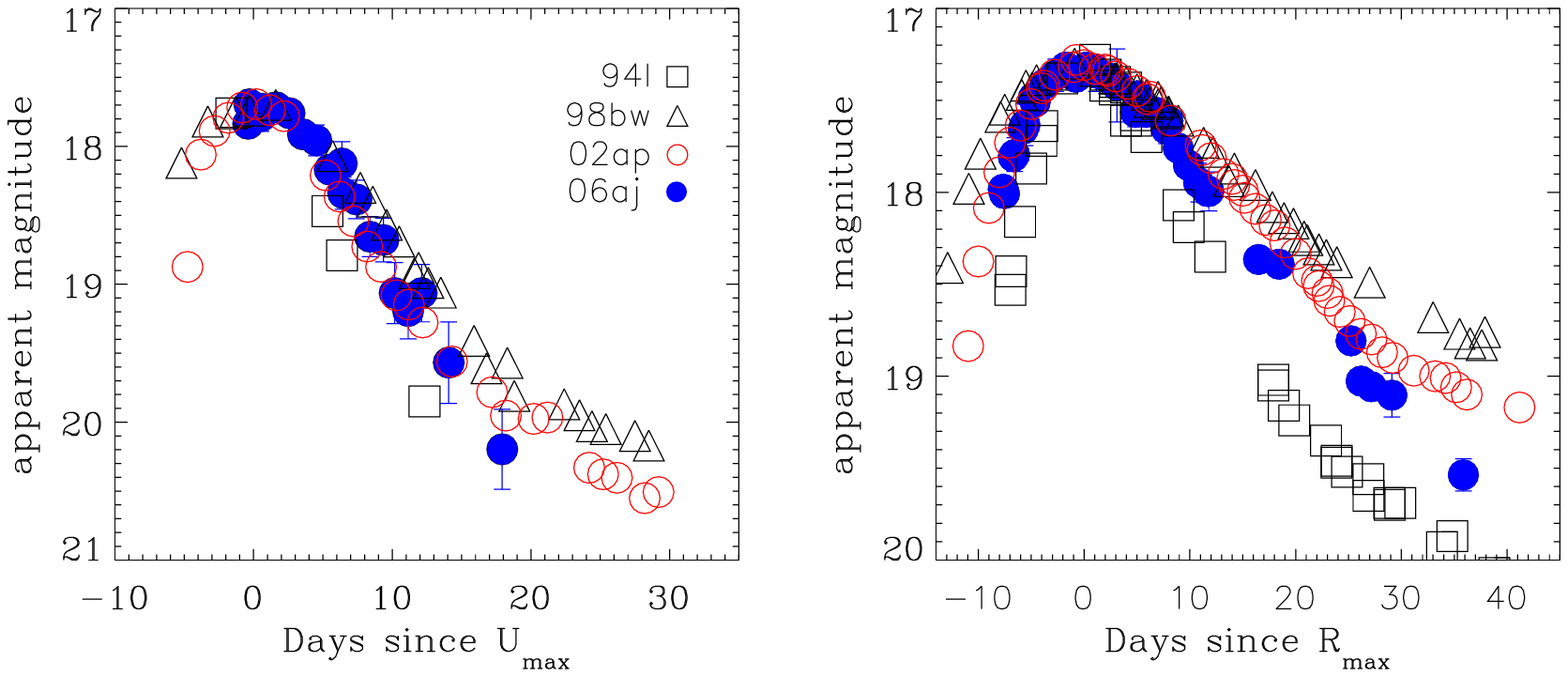}}
\caption{}{Comparison of $U$- and $R$-band lightcurves for SN\,2006aj and three other Type Ic SNe. SN\,1998bw is seen to evolve quite a bit slower, while SN\,1994I is clearly faster. The best match is with SN\,2002ap.
The lightcurves have been corrected for time dilation and matched at date of peak and at maximum brightness. In this plot we have also corrected SN\,2006aj for the host galaxy contribution. Note that the final $U$-band datapoint is uncertain due to a large relative host extinction correction.
}
\label{f:lightcurvescompare}
\end{flushleft}
\end{figure*}

\begin{figure*}[h]
\begin{flushleft}
{\includegraphics[width=0.5\textwidth,angle=0,clip]{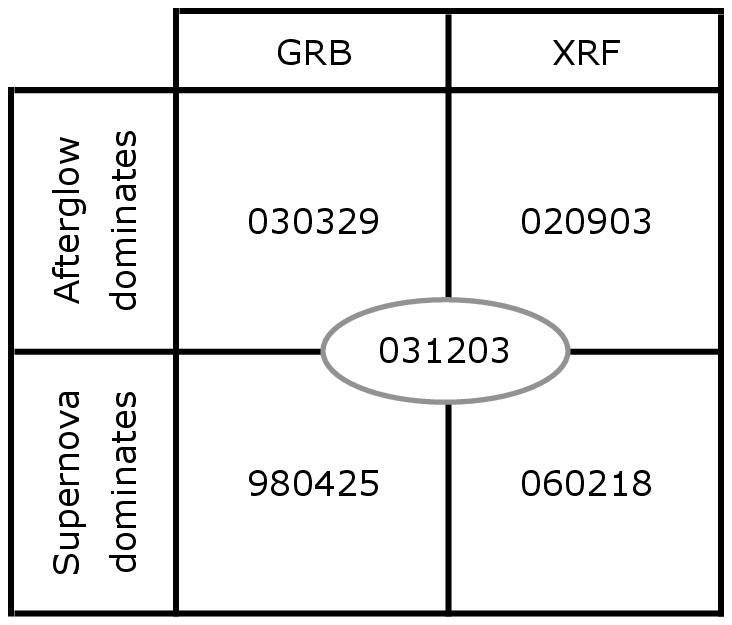}}
\caption{}{
Four-field diagram representing different varieties of Supernova-GRBs.
There are differences in high-energy properties defining the peak energy of the burst, as well as in the optical afterglow appearance. XRF\,060218 / SN\,2006aj fills in the lower right field of this diagram, as an XRF with supernova light dominating the early optical transient.
}
\label{f:xrfs}
\end{flushleft}
\end{figure*}

From Fig.~\ref{f:xrfs} we see that it may be difficult to reconcile these diverse observations by a simple geometric scenario. In this four-field diagram we have divided bursts into XRFs and GRBs. We have also divided them according to the dominating component in the optical lightcurve; supernova or afterglow. We have indicated the spectroscopically confirmed SN-GRBs. The upper left corner is represented by GRB\,030329 where SN\,2003dh was not apparent until after a week. XRF\,020903 was dominated by an afterglow until the late supernova bump, and occupies the upper right field. The same applies to XRF\,030723. The lower left box is represented by SN\,1998bw and GRB\,980425, although the peak energy was not very high. Finally, XRF\,060218 now fills in the lower right field in this diagram. It has a very low E$_{\rm peak}$ and shows supernova signatures already from the very early photometry \citep{campana06} and spectroscopy \citep[][and this work Fig.~\ref{f:spectra}]{modjaz06,mirabal06}.

A one-parameter explanation such as an on- vs. off-axis picture would have problem to explain all the combinations in Fig.~\ref{f:xrfs}. It seems that (SN)-XRFs can come both with and without a conspicious afterglow, and the afterglow can moreover behave quite differently (flat early lightcurve in XRF\,030723 vs. constant decay in XRF\,050824). A larger sample of SN-GRBs will be needed to unveil whether we observe different classes of objects, or simply a continuum of burst properties.

\citet{thomsen04} actually predicted Swift to detect
a significant population of faint bursts and
hence allow the study of core-collapse SNe at much
earlier times than had been previously possible, and indicated that this would 
have a substantial impact on SN research. The discovery of the first nearby Swift GRB-SN substantiates this prediction.

\begin{acknowledgements}
This work was 
done at the Dark Cosmology Centre funded by The
Danish National Research Foundation.
JS also 
acknowledge support from Danmarks Nationalbank and from the Anna-Greta and 
Holger Crafoord fund.  AOJ acknowledges support from the Norwegian Research
Council.
Some of the data presented here have been taken using ALFOSC, which is owned by the Instituto de Astrofisica de Andalucia (IAA) and operated at the Nordic Optical Telescope under agreement between IAA and the NBI of the Astronomical Observatory of Copenhagen.% 
\end{acknowledgements}

%%% table 1 is the log of observations and the magnitudes

%\begin{deluxetable}{llccccl}
\begin{longtable}{llccccl}
\caption{\label{t:log} Log of observations and photometry of supernova 2006aj.}\\
%\tablehead{
\hline\hline
Date & $\Delta$t & Pass band & Exptime & Magnitude &
Magnitude Error & Telescope \\
(UT) & (days) & & (s) & & ($1\sigma)$ & \\
\hline
\endfirsthead
\caption{continued.}\\
\hline\hline
Date & $\Delta$t & Pass band & Exptime & Magnitude &
Magnitude Error & Telescope \\
(UT) & (days) & & (s) & & ($1\sigma)$ & \\
\hline
\endhead
\hline
\endfoot
%\startdata
\hline
Feb 23.932 &   5.783 &   U &  900 &  17.74 & 0.06 & NOT\\
Feb 24.006 &   5.857 &   U &  120 &  17.43 & 0.30 & D1.5m\\
Feb 24.009 &   5.860 &   U &  200 &  17.33 & 0.16 & D1.5m\\
Feb 24.012 &   5.863 &   U &  200 &  17.55 & 0.07 & D1.5m\\
Feb 24.016 &   5.867 &   U &  200 &  17.61 & 0.06 & D1.5m\\
Feb 24.022 &   5.873 &   U &  600 &  17.66 & 0.04 & D1.5m\\
Feb 24.030 &   5.881 &   U &  600 &  17.60 & 0.04 & D1.5m\\
Feb 24.898 &   6.749 &   U &  300 &  17.69 & 0.10 & NOT\\
Feb 25.009 &   6.860 &   U &  200 &  17.62 & 0.09 & D1.5m\\
Feb 25.014 &   6.865 &   U &  400 &  17.67 & 0.06 & D1.5m\\
Feb 25.021 &   6.872 &   U &  600 &  17.64 & 0.04 & D1.5m\\
Feb 25.029 &   6.880 &   U &  600 &  17.63 & 0.05 & D1.5m\\
Feb 26.006 &   7.857 &   U &  200 &  17.65 & 0.24 & D1.5m\\
Feb 26.011 &   7.862 &   U &  200 &  17.51 & 0.12 & D1.5m\\
Feb 26.016 &   7.867 &   U &  600 &  17.66 & 0.03 & D1.5m\\
Feb 26.024 &   7.875 &   U &  600 &  17.71 & 0.03 & D1.5m\\
Feb 27.008 &   8.859 &   U &  200 &  17.35 & 0.17 & D1.5m\\
Feb 27.011 &   8.862 &   U &  200 &  17.61 & 0.10 & D1.5m\\
Feb 27.017 &   8.868 &   U &  600 &  17.69 & 0.05 & D1.5m\\
Feb 27.025 &   8.876 &   U &  600 &  17.72 & 0.05 & D1.5m\\
Feb 28.007 &   9.858 &   U &  200 &  17.57 & 0.16 & D1.5m\\
Feb 28.010 &   9.861 &   U &  200 &  17.79 & 0.10 & D1.5m\\
Feb 28.016 &   9.867 &   U &  600 &  17.77 & 0.08 & D1.5m\\
Feb 28.024 &   9.875 &   U &  600 &  17.87 & 0.04 & D1.5m\\
Mar  1.012 &  10.863 &   U &  600 &  17.85 & 0.10 & D1.5m\\
Mar  2.003 &  11.854 &   U &  200 &  18.48 & 0.24 & D1.5m\\
Mar  2.007 &  11.858 &   U &  200 &  18.16 & 0.25 & D1.5m\\
Mar  2.012 &  11.863 &   U &  600 &  18.04 & 0.08 & D1.5m\\
Mar  2.914 &  12.765 &   U &  300 &  18.00 & 0.14 & NOT\\
Mar  3.003 &  12.854 &   U &  200 &  18.06 & 0.18 & D1.5m\\
Mar  3.009 &  12.860 &   U &  600 &  18.19 & 0.10 & D1.5m\\
Mar  4.005 &  13.856 &   U &  200 &  18.01 & 0.15 & D1.5m\\
Mar  4.010 &  13.861 &   U &  600 &  18.45 & 0.08 & D1.5m\\
Mar  5.005 &  14.856 &   U &  200 &  18.31 & 0.18 & D1.5m\\
Mar  5.011 &  14.862 &   U &  600 &  18.46 & 0.12 & D1.5m\\
Mar  6.006 &  15.857 &   U &  200 &  18.57 & 0.20 & D1.5m\\
Mar  6.012 &  15.863 &   U &  600 &  18.48 & 0.13 & D1.5m\\
Mar  6.851 &  16.702 &   U &  900 &  18.79 & 0.17 & NOT\\
Mar  7.854 &  17.705 &   U &  900 &  18.89 & 0.15 & NOT\\
Mar  8.866 &  18.717 &   U &  900 &  18.79 & 0.16 & NOT\\
Mar 10.878 &  20.729 &   U & 1500 &  19.16 & 0.20 & NOT\\
Mar 14.864 &  24.715 &   U & 2400 &  19.55 & 0.15 & NOT\\
\hline
Feb 21.013 &   2.864 &   V &  120 &  18.22  & 0.05 & D1.5m\\
Feb 21.015 &   2.866 &   V &  120 &  18.16  & 0.08 & D1.5m\\
Feb 21.017 &   2.868 &   V &  120 &  18.21  & 0.03 & D1.5m\\
Feb 21.034 &   2.885 &   V &  300 &  18.21  & 0.03 & D1.5m\\
Feb 21.051 &   2.902 &   V &  600 &  18.17  & 0.03 & D1.5m\\
Feb 22.020 &   3.871 &   V &  120 &  18.01  & 0.03 & D1.5m\\
Feb 22.022 &   3.873 &   V &  120 &  18.02  & 0.05 & D1.5m\\
Feb 22.040 &   3.891 &   V &  300 &  18.01  & 0.03 & D1.5m\\
Feb 22.045 &   3.896 &   V &  300 &  18.00  & 0.03 & D1.5m\\
Feb 22.049 &   3.900 &   V &  300 &  18.03  & 0.03 & D1.5m\\
Feb 23.027 &   4.878 &   V &  300 &  17.84  & 0.05 & D1.5m\\
Feb 23.031 &   4.882 &   V &  300 &  17.87  & 0.02 & D1.5m\\
Feb 23.046 &   4.897 &   V &  300 &  17.84  & 0.03 & D1.5m\\
Feb 23.050 &   4.901 &   V &  300 &  17.85  & 0.04 & D1.5m\\
Feb 24.041 &   5.892 &   V &  300 &  17.71  & 0.04 & D1.5m\\
Feb 25.041 &   6.892 &   V &  300 &  17.61  & 0.03 & D1.5m\\
Feb 26.031 &   7.882 &   V &  300 &  17.53  & 0.03 & D1.5m\\
Feb 27.032 &   8.883 &   V &  300 &  17.50  & 0.04 & D1.5m\\
Feb 28.031 &   9.882 &   V &  300 &  17.49  & 0.04 & D1.5m\\
Mar  1.018 &  10.869 &   V &  200 &  17.50  & 0.08 & D1.5m\\
Mar  2.019 &  11.870 &   V &  200 &  17.50  & 0.03 & D1.5m\\
Mar  2.924 &  12.775 &   V &  300 &  17.52  & 0.04 & NOT\\
Mar  3.016 &  12.867 &   V &  200 &  17.54  & 0.05 & D1.5m\\
Mar  4.017 &  13.868 &   V &  200 &  17.61  & 0.03 & D1.5m\\
Mar  5.018 &  14.869 &   V &  200 &  17.62  & 0.15 & D1.5m\\
Mar  6.863 &  16.714 &   V &  300 &  17.77  & 0.10 & NOT\\
Mar  7.002 &  16.853 &   V &  300 &  17.74  & 0.04 & D1.5m\\
Mar  7.012 &  16.863 &   V &  300 &  17.81  & 0.02 & D1.5m\\
Mar  7.863 &  17.714 &   V &  300 &  17.85  & 0.08 & NOT\\
Mar  8.002 &  17.853 &   V &  300 &  17.92  & 0.07 & D1.5m\\
Mar  8.006 &  17.857 &   V &  300 &  17.84  & 0.07 & D1.5m\\
Mar  8.874 &  18.725 &   V &  300 &  18.00  & 0.09 & NOT\\
Mar  9.008 &  18.859 &   V &  450 &  17.94  & 0.04 & D1.5m\\
Mar  9.837 &  19.688 &   V &  300 &  18.04  & 0.11 & NOT\\
Mar 11.004 &  20.855 &   V &  300 &  18.12  & 0.07 & D1.5m\\
Mar 12.000 &  21.851 &   V &  200 &  18.20  & 0.07 & D1.5m\\
Mar 12.999 &  22.850 &   V &  100 &  18.24  & 0.17 & D1.5m\\
Mar 13.859 &  23.710 &   V &  900 &  18.26  & 0.06 & NOT\\
Mar 13.995 &  23.846 &   V &  100 &  18.29  & 0.09 & D1.5m\\
Mar 15.992 &  25.843 &   V &  180 &  18.34  & 0.12 & D1.5m\\
Mar 18.840 & 28.691  &   V &  900 &  18.65  & 0.07 & NOT\\
Mar 20.852 & 30.703  &   V &  900 &  18.78  & 0.06  & NOT\\
\hline
Feb 21.021 &   2.872 &   R &  200 &   18.00 & 0.02 & D1.5m\\
Feb 21.025 &   2.876 &   R &  300 &   17.97 & 0.09 & D1.5m\\
Feb 21.029 &   2.880 &   R &  300 &   17.99 & 0.09 & D1.5m\\
Feb 21.041 &   2.892 &   R &  600 &   17.99 & 0.07 & D1.5m\\
Feb 21.057 &   2.908 &   R &  150 &   18.00 & 0.08 & D1.5m\\
Feb 21.915 &   3.766 &   R &  900 &   17.79 & 0.03 & NOT\\
Feb 22.026 &   3.877 &   R &  300 &   17.80 & 0.04 & D1.5m\\
Feb 22.031 &   3.882 &   R &  300 &   17.82 & 0.06 & D1.5m\\
Feb 22.054 &   3.905 &   R &  300 &   17.81 & 0.08 & D1.5m\\
Feb 22.898 &   4.749 &   R &  900 &   17.64 & 0.03 & NOT\\
Feb 23.016 &   4.867 &   R &  200 &   17.62 & 0.07 & D1.5m\\
Feb 23.019 &   4.870 &   R &  200 &   17.64 & 0.08 & D1.5m\\
Feb 23.023 &   4.874 &   R &  200 &   17.63 & 0.09 & D1.5m\\
Feb 23.036 &   4.887 &   R &  300 &   17.61 & 0.08 & D1.5m\\
Feb 23.041 &   4.892 &   R &  300 &   17.64 & 0.09 & D1.5m\\
Feb 23.879 &   5.730 &   R &  250 &   17.51 & 0.04 & NOT\\
Feb 24.036 &   5.887 &   R &  300 &   17.49 & 0.10 & D1.5m\\
Feb 24.854 &   6.705 &   R &  300 &   17.36 & 0.04 & NOT\\
Feb 25.036 &   6.887 &   R &  300 &   17.38 & 0.05 & D1.5m\\
Feb 26.036 &   7.887 &   R &  300 &   17.29 & 0.03 & D1.5m\\
Feb 27.037 &   8.888 &   R &  300 &   17.25 & 0.08 & D1.5m\\
Feb 28.035 &   9.886 &   R &  300 &   17.21 & 0.02 & D1.5m\\
Mar  1.023 &  10.874 &   R &  200 &   17.26 & 0.04 & D1.5m\\
Mar  2.022 &  11.873 &   R &  200 &   17.21 & 0.03 & D1.5m\\
Mar  2.849 &  12.700 &   R &  300 &   17.22 & 0.03 & NOT\\
Mar  3.020 &  12.871 &   R &  200 &   17.24 & 0.09 & D1.5m\\
Mar  4.020 &  13.871 &   R &  200 &   17.26 & 0.08 & D1.5m\\
Mar  5.022 &  14.873 &   R &  200 &   17.30 & 0.18 & D1.5m\\
Mar  6.843 &  16.694 &   R &  300 &   17.43 & 0.07 & NOT\\
Mar  7.016 &  16.867 &   R &  300 &   17.34 & 0.06 & D1.5m\\
Mar  7.020 &  16.871 &   R &  200 &   17.35 & 0.03 & D1.5m\\
Mar  7.867 &  17.718 &   R &  300 &   17.43 & 0.06 & NOT\\
Mar  8.012 &  17.863 &   R &  300 &   17.40 & 0.02 & D1.5m\\
Mar  8.017 &  17.868 &   R &  300 &   17.43 & 0.05 & D1.5m\\
Mar  8.848 &  18.699 &   R & 1500 &   17.43 & 0.03 & NOT\\
Mar  9.841 &  19.692 &   R &  300 &   17.51 & 0.07 & NOT\\
Mar 10.010 &  19.861 &   R &  400 &   17.47 & 0.07 & D1.5m\\
Mar 11.009 &  20.860 &   R &  300 &   17.60 & 0.03 & D1.5m\\
Mar 12.005 &  21.856 &   R &  400 &   17.68 & 0.02 & D1.5m\\
Mar 13.001 &  22.852 &   R &  150 &   17.76 & 0.09 & D1.5m\\
Mar 13.872 &  23.723 &   R &  900 &   17.80 & 0.04 & NOT\\
Mar 13.998 &  23.849 &   R &  100 &   17.79 & 0.08 & D1.5m\\
Mar 14.000 &  23.851 &   R &  100 &   17.81 & 0.06 & D1.5m\\
Mar 18.860 & 28.711  &   R &  900 &       18.10 & 0.04 & NOT\\
Mar 20.866 & 30.717  &   R  & 900 &       18.12 & 0.04 & NOT\\
Mar 27.889 & 37.740  &   R  & 1800&  18.43& 0.04 & NOT\\
Mar 28.874 & 38.725  &   R  & 1800&  18.58& 0.03 & NOT\\
Mar 29.874 &  39.725 &   R  & 2100&  18.60&  0.03& NOT\\
Mar 31.884 & 41.735  &   R  & 1800&  18.63& 0.08& NOT\\
Apr 07.858 & 48.709  &   R  & 1800&   18.89& 0.05&NOT\\

\hline  
\end{longtable}

%%% table 4 is the spectral lines

\begin{table}% {lcccc}
\caption{Strong emission lines. 
\label{t:lines}}
\centering   
\begin{tabular}{l c c c c}
%\tablehead{
\hline\hline  
ID & Rest Wavelength & Observed Wavelength & Flux & Redshift  \\    
   &  (\AA) &           (\AA) & (10$^{-16}$ erg s$^{-1}$ cm$^{-2})$ & \\    
\hline\hline  
%\startdata
\hline
O II    & 3727.42  &     3853.42  &    16  &  0.0338 \\
H$\beta$    & 4861.33 &    5021.00  &   11    &0.0328 \\
O III    & 4958.91  &    5121.76   &    16 &   0.0328 \\
O III   & 5006.84  &    5171.18   &    46 &   0.0328 \\
H$\alpha$   & 6563.00 &    6778.14       &20    &0.0328   \\
\hline  
\end{tabular}
\end{table}

\begin{table}%{lccc}
\caption{Final corrected light curve estimates.
\label{t:summary}}
\centering   
\begin{tabular}{l c c c}
\hline\hline  
 & $U$ & $V$ & $R$ \\
\hline
%\startdata
Rise time (days) & 6.6 & 10.0 & 11.0 \\
$\Delta$m$_{15}$ (mag)  & 2.0 & 1.1 & 0.90 \\
Abs. Mag & $-18.95$ & $-18.65$ & $-18.68$ \\
\hline                                   %inserts single line
\end{tabular}
\end{table}

%%%%%%%%%%%%%%%  Figures   
%%%%%%  Fig 1 LCs

%%% Fig 2 is the spectra

\begin{thebibliography}{}

%\bibitem[Allen(2000)]{allen00}
%Allen, C. W. 2000, Allen's Astrophysical Quantities, 4th edition 2000,
%ed. A. N. Cox

\bibitem[Bersier et al.(2006)]{bersier06}
Bersier, D., Fruchter, A.~S., Strogler, L.-G., et al. 2006, astro-ph/0602163

\bibitem[Bohlin et al.(1995)]{bohlin95}
Bohlin, R.~C., Colina, L., \& Finley, D.~S. 1995, AJ, 110, 1316


\bibitem[Bolzonella et al.(2000)]{bolzonella00}
Bolzonella, M., Miralles, J.-M.,  \& Pell{\o}, R. 2000, A\&A, 363, 476

\bibitem[Calzetti et al.(2000)]{calzetti00}
Calzetti, D., Armus, L., Bohlin, R.~C., et  al. 2000,  ApJ,  533, 682


\bibitem[Bertin \& Arnouts(1996)]{bertin96}
Bertin, E., \&  Arnouts, S. 1996, A\&AS, 117, 393	


\bibitem[Campana et al.(2006)]{campana06}
Campana, S., Mangano, V., Blusin, A.~J., et al. 2006, submitted to Nature, astro-ph/0603279

\bibitem[Cool et al.(2006)]{coolgcn4777}
Cool, R.~J., Eisenstein, D.~J., Hogg, D.~W., et al. 2006, GCN, 4777

\bibitem[Cusumano et al.(2006)]{cusumanogcn4775}
Cusumano, G., Barthelmy, S., Gehrels, N., et al. 2006, GCN, 4775

\bibitem[Fitzpatrick(1986)]{fitzpatrick86}
Fitzpatrick, E.~L. 1986, AJ 92, 1068


%\bibitem[Fukugita et al.(1995)]{fukugita95}
%Fukugita, M., Shimasaku, K., \& Ichikawa, T. 1995, \pasp, 107, 945


\bibitem[Fukugita et al.(1996)]{fukugita96}
Fukugita, M., Ichikawa, T., Gunn, J.~E., Doi, M., Shimasaku, K., \& Schneider, D. P. 1996, AJ, 111, 1748


\bibitem[Foley et al.(2003)]{foley03}
Foley, R.~J., Papenkova, M.~S., Swift, B.~J., et al.
2003, PASP, 115, 1220	


\bibitem[Fynbo et al.(2004)]{fynbo04}
Fynbo, J., Sollerman, J., Hjorth, J., et al. 2004, ApJ, 609, 962


\bibitem[Galama et al.(1998)]{galama98}
Galama, T. J., Vreeswijk, P. M., van Paradijs, J., et al. 1998, Nature, 395, 670


\bibitem[Gal-Yam et al.(2002)]{galyam02}
Gal-Yam, A., Ofek, E.~O., \&  Shemmer, O. 2002, 
MNRAS, 332, 73


\bibitem[Gehrels et al.(2004)]{gehrels04}
Gehrels, N., Chincarini, G., Giommi, P., et al.
2004, ApJ, 611, 1005	


\bibitem[Gorosabel et al.(2005)]{gorosabel05}
Gorosabel, J., Perez-Ramirez, D., Sollerman, J., 
et al. 2005, A\&A, 444, 711

\bibitem[Guenther et al.(2006)]{guenthergcn4863} 
Guenther, E.~W., Klose, S., Vreeswijk, P., et al. 2006, GCN, 4863

\bibitem[Hicken et al.(2006)]{hicken06gcn4898}
Hicken, M., Modjaz, M., Challis, P., et al. 2006, GCN, 4898

\bibitem[Hjorth et al.(2003)]{hjorth03}
Hjorth, J., Sollerman, J., M\o ller, P., et al. 2003, \nat, 423, 847

%\bibitem[Hjorth et al.(2003b)]{hjorth03b}
%Hjorth, J., M\o ller, P., Gorosabel, J., et al., 2003b, \apj, 597, 699

\bibitem[Hoeflich et al.(1999)]{hoeflich99}
H\"oflich, P.,  Wheeler, J.~C., \&  Wang, L. 1999,
ApJ, 521, 179

%\bibitem[Jakobsson et al.(2004)]{jakobsson04}
%Jakobsson, P., Hjorth, J., Fynbo, J.~P.~U. 2004,  A\&A, 425, 785

\bibitem[Jakobsson et al.(2006)]{jakobsson06}
Jakobsson, P., Levan, A., Fynbo, J.~P.~U., et al.
2006, A\&A, 447, 897	

\bibitem[Jester et al.(2005)]{jester05}
Jester, S., Schneider, D.~P., Richards, G.~T.,
et al. 2005, AJ, 130, 873


\bibitem[Kennea et al.(2006)]{kenneagcn4776}
Kennea, J.~A., Burrows, D.~N., Cusumano, G., et al. 2006, GCN, 4776

\bibitem[Kennicutt(1998)]{kennicutt98}
Kennicutt, R. C. 1998, ARA\&A, 36, 189


\bibitem[Kewley \& Dopita(2002)]{kewley02}
Kewley, L. J., \& Dopita, M. A.
2002, ApJS, 142, 35


\bibitem[Landolt(1992)]{landolt}
Landolt, A.~U. 1992, AJ, 104, 340


\bibitem[Malesani et al.(2004)]{malesani04} 
Malesani, D., Tagliaferri, G., Chincarini, G., et al. 
2004, ApJ, 609, 5	

\bibitem[Matheson et al.(2003)]{matheson03} 
Matheson, T., Garnavich, P.~M., Stanek, K.~Z.,
et al.\  2003, \apj, 599, 394 


\bibitem[Masetti et al.(2006)]{masettigcn4803} 
Masetti, N., Palazzi, E.,  Pian, E., et al. 2006,
GCN, 4803

\bibitem[Mazzali et al.(2002)]{mazzali02} 
Mazzali, P.~A., Deng, J., Maeda, K., et al.
2002, ApJ, 572, 61	


\bibitem[Miller \& Scalo(1979)]{miller79}
Miller,  G.~E., \& Scalo, J.~M. 1979, ApJS, 41, 513


\bibitem[Mirabal et al.(2006)]{mirabalgcn4792} 
Mirabal, M., Halpern, J.~P., et al. 2006, GCN, 4792

\bibitem[Mirabal et al.(2006)]{mirabal06} 
Mirabal, M., Halpern, J.~P., An, D., Thorstensen, J.~R., \& Terndrup, D.~M.
2006, astro-ph/0603686


\bibitem[Modjaz et al.(2006)]{modjaz06} 
Modjaz, M., Stanek, K.~Z., Garnavich, P.~M., et al. 2006, astro-ph/0603377

\bibitem[Munari \& Zwitter(1997)]{munari97} 
Munari, U., \& Zwitter, T.
1997, A\&A, 318, 269	


\bibitem[Pandey et al.(2003)]{pandey03} 
Pandey, S. B., Anupama, G. C., Sagar, R., Bhattacharya, D., Sahu, D. K., 
\& Pandey, J. C.	2003, MNRAS, 340, 375	


\bibitem[Patat et al.(2001)]{patat01}
Patat, F., Cappellaro, E., Danziger, J., et al. 2001, \apj, 555, 900

\bibitem[Pei(1992)]{pei92}
Pei, Y.~C. 1992, \apj, 395, 130 

\bibitem[Pian et al.(2006)]{pian06}
Pian, E., Mazzali, P., Masetti, N., et al. 2006, astro-ph/0603530

\bibitem[Prevot et al.(1984)]{prevot84}
Prevot, M.~L., Lequeux, J., Prevot, L., Maurice, E., \& Rocca-Volmerange, B.
1984, A\&A, 132, 389

\bibitem[Ramirez-Ruiz et al.(2005)]{ramirez05}
Ramirez-Ruiz, E., 
Granot, J., Kouveliotou, C., Woosley, S.~E., Patel, S.~K., \& Mazzali, 
P.~A.\ 2005, \apjl, 625, L91 
 

\bibitem[Richardson et al.(2006)]{richardson06}
Richardson, D., Branch, D., \& Baron, E. 2006, astro-ph/0601136

\bibitem[Richmond et al.(1996)]{richmond96}
Richmond, M. W., van Dyk, S. D., Ho, W. et al. 1996, AJ, 111, 327


\bibitem[Salpeter(1955)]{salpeter55}
Salpeter, E.~E.  1955, ApJ, 121, 161 

\bibitem[Scalo(1986)]{scalo86}
Scalo, J.~M. 1986, Fundam. Cosmic Phys. 11, 1 


\bibitem[Schlegel et al.(1998)]{schlegel98} 
Schlegel, D. J., Finkbeiner, D. P., \&  Davis, M. 1998, \apj, 500, 525.

\bibitem[Seaton(1979)]{seaton79}
Seaton, M.~J. 1979, MNRAS, 187, 73 


\bibitem[Soderberg et al.(2005)]{soderberg05} 
Soderberg, A.~M., Kulkarni, S.~R., Fox, D.~B.,  et  al.\ 2005, \apj, 627, 877 


\bibitem[Soderberg et al.(2006a)]{soderberggcn4794} 
Soderberg, A.~M., Frail, D., et al. 2006, GCN, 4794

\bibitem[Soderberg et al.(2006b)]{soderbergiau} 
Soderberg, A., Berger, E., \& Schmidt, B. 2006, IAUC, 8674


\bibitem[Sollerman et al.(2000)]{sollerman00} 
Sollerman, J., Kozma, C.,  Fransson, C., Leibundgut, B., Lundqvist, P.,
Ryde, F., \& Woudt, P. 2000, ApJL, 537, 127

%\bibitem[Sollerman et al.(2002)]{sollerman02} 
%Sollerman, J., Holland, S., Challi, P., et al. 2000 A\&A


\bibitem[Sollerman et al.(2005a)]{sollerman05a} 
Sollerman, J., Cox, N., Mattila, S., et al.
2005a, A\&A, 429, 559

\bibitem[Sollerman et al.(2005b)]{sollerman05b} 
Sollerman, J., {\"O}stlin, G., Fynbo, J.~P.~U., Hjorth, J., Fruchter, A., \& Pedersen, K.\ 2005b, New Astronomy, 11, 103 


\bibitem[Sollerman et al.(2006)]{sollerman06} 
Sollerman, J., Fynbo, J.~P.~U., et al. 2006, in prep.

\bibitem[Stanek et al.(2003)]{stanek03}
Stanek, K. Z., Matheson, T., Garnavich, P. M., et al. 2003, \apj, 591, L17

\bibitem[Stetson(1987)]{stetson87}
Stetson, P.~B. 1987, PASP, 99, 191	


\bibitem[Stritzinger et al.(2002)]{stritzinger02}
Stritzinger, M., Hamuy, M., Suntzeff, N.~B., et al.
2002, AJ, 124, 2100	


\bibitem[Stritzinger et al.(2006)]{stritzinger05}
Stritzinger, M., Leibundgut, B., Walch, S., \& Contardo, G. 2006, 
A\&A, in press, astro-ph/0506415

\bibitem[Thomsen et al.(2004)]{thomsen04}
Thomsen, B., Hjorth, J., Watson, D., et al. 2004, A\&A, 419, 21

\bibitem[Watson et al.(2004)]{watson04}
Watson, D., Hjorth, J., Levan, A., et al.
2004, ApJ, 605, 101	

\bibitem[Woosley et al.(1999)]{woosley99}
Woosley, S.~E., Eastman, R.~G., \& Schmidt, B.~P. 1999, ApJ, 516, 788 

\end{thebibliography}
\end{document}